\documentclass[aps,prb,twocolumn,showpacs,groupedaddress]{revtex4}
\usepackage{graphicx}
\begin{document}

\newcommand{\azu}{Cu$_3$(CO$_3$)$_2$(OH)$_2$}


\title{Magnetoelastic and structural properties of azurite Cu$_3$(CO$_3$)$_2$(OH)$_2$ from neutron scattering and muon spin rotation}
\author{M.C.R. Gibson$^{1,2}$, K.C. Rule$^1$, A.U.B. Wolter$^3$, J.-U. Hoffmann$^1$, O. Prokhnenko$^1$, D.A. Tennant$^{1,2}$, S. Gerischer$^1$, M. Kraken$^4$, F.J. Litterst$^4$, S. S\"ullow$^4$, J. Schreuer$^5$, H. Luetkens$^6$, A. Br\"{u}hl$^{7}$, B. Wolf$^{7}$, M. Lang$^{7}$, }
\affiliation{$^1$Helmholtz-Zentrum Berlin, Berlin, Germany\\
$^2$Institut f\"ur Festk\"orperphysik, TU Berlin, Berlin, Germany\\
$^3$Institut f\"ur Festk\"{o}rper- und Werkstoffforschung, Dresden, Germany\\
$^4$Institut f\"{u}r Physik der Kondensierten Materie, TU Braunschweig, Braunschweig, Germany\\
$^5$Ruhr-Universit\"at Bochum, Bochum, Germany\\
$^6$Paul-Scherrer-Institut, Villigen, Switzerland\\
$^7$Physikalisches Institut, J.W. Goethe Universit\"{a}t, Frankfurt(M), Germany}

\date{\today}

\begin{abstract}
Azurite, \azu, has been considered an ideal example of a one-dimensional (1D) diamond chain antiferromagnet.  
Early studies of this material imply the presence of an ordered antiferromagnetic phase below $T_N \sim 1.9$ K  while magnetization measurements have revealed a 1/3 magnetization plateau.  
Until now, no corroborating neutron scattering results have been published to confirm the ordered magnetic moment structure.  
We present recent neutron diffraction results which reveal the presence of commensurate magnetic order in azurite which coexists with significant magnetoelastic strain.  The latter of these effects may indicate the presence of spin frustration in zero applied magnetic field. Muon spin rotation $\mu$SR reveals an onset of short-range order below 3K and confirms long-range order below $T_N$.
\end{abstract}


\maketitle

Despite the burgeoning volume of data on the natural mineral azurite and its magnetic properties, the precise details of its spin-ordered state and the microscopic exchange couplings remain contentious issues. Recently, strong magnetoelastic coupling has been observed in azurite indicating significant interdependence of structural and magnetic degrees of freedom\cite{Fabris}. 
This effect is reminiscent of the spin-Peierls cuprate system CuGeO$_3$\cite{hase} in which the distortion is thought to be due to competing antiferromagnetic (AFM) interactions.  In contrast, the structural distortion in azurite coincides with a transition to a three-dimensionally ordered AFM state.

Azurite, \azu\ is a \textit{quasi}-1D system and the first realization of the distorted diamond chain \cite{kikuchi1,kikuchi2}. 
Here, Cu$^{2+}$ ions form a triangular arrangement of spin $S = 1/2$ moments which are arranged as chains propagating along the crystallographic $b$-axis. 
The diamond chain model has been extensively studied \cite{takano,okamoto1,tonegawa,honecker,okamoto2}, with the distortion implied by three inequivalent exchange couplings, $J_1 \neq J_2 \neq J_3$.
This model affords a host of exotic phases and quantum phase transitions, including exotic dimer phases \cite{okamoto1,tonegawa,okamoto2} or $M = 1/3$ fractionalisation \cite{mueller,oshikawa}.

Azurite has a monoclinic crystal structure (space group $\it P2_1/c$) with room temperature lattice parameters $a$ = 5.01 \AA , $b$ = 5.85 \AA , $c$ = 10.3 \AA , and a monoclinic angle, $\beta = 92.4^{\circ}$ \cite{gattow,zigan,belokoneva}.  
An NMR study initially revealed the AFM phase transition at $T_N = 1.86$K\cite{spence} while  magnetization measurements have  indicated the presence of a distinct 1/3 magnetization plateau \cite{kikuchi1,kikuchi2}.
Long range magnetic order was also confirmed by ESR measurements revealing AFM resonance modes below 1.9K \cite{okubo}.
Additional studies, including muon spin resonance, indicated that the magnetic phase transition at $T_N$ is of second order \cite{forstat,garber,love,kikuchi3}.

Recently, there has been some controversy over the effective dimensionality of azurite.
\textit{Ab initio} density functional calculations have suggested that there is sizeable magnetic exchange between the diamond chains \cite{whangbo}.
Inelastic neutron scattering measurements exhibit dispersive magnetic excitations along the chain\cite{rule}, indicating that significant magnetic exchange is present in this direction.
Subsequent measurements of the dynamical spin correlations in both directions perpendicular to the chain have revealed no dispersion\cite{future}. 
Thus the dynamics appear well defined by an effective one-dimensional model.

Now, while the study of spin dynamics can yield estimates of effective magnetic couplings within azurite, a detailed study of the static spin arrangement reveals much about the strength of the microscopic exchange interactions and spin anisotropy.  This motivated the current survey of the magnetic ordering transition in azurite.
   
Here we present measurements from both single crystal and polycrystalline samples using a variety of neutron diffraction methods as well as $\mu$SR measurements. 
We present the first observation of magnetic order from which we have determined the magnetic propagation vector of the commensurate structure. 
Importantly, we also observed significant lattice distortions in the single crystalline material indicating a strong magneto-elastic coupling in the lattice. This lattice strain is consistent with the results of high-resolution thermal expansion measurements to be published in more detail elsewhere \cite{bruhl}. 

The samples used in this study were derived from the same high-quality single crystal of \azu \cite{rule}.  
A cube was cut with length $\sim$ 5mm for the single crystal neutron measurements.  A smaller crystallite, oriented along the $b$-axis, was used for the thermal expansion measurements. For this, the resolution was greater than $\Delta l/l = 10^{-10}$, where $l$ is the sample length.
Smaller crystallites were crushed thoroughly for the powder diffraction studies. 
The powder experiments were performed using the E9 diffractometer at the Helmholtz Zentrum Berlin (HZB) (neutron wavelengh $\lambda = 1.798$\AA ) at temperatures $T$ = 5 and 1.28 K.
Single crystal neutron diffraction has been carried out using the E1 triple-axis spectrometer (TAS) at HZB (neutron wavelength $\lambda = 2.43$\AA ).  
The sample was mounted with a horizontal $a^* - b^*$ scattering plane within a $^{3}$He insert to ensure good thermal stability below $T_N \sim 1.9K$.
Further single crystal neutron measurements were performed on the flat-cone diffractometer, E2, also at HZB at 2 temperatures, 1.4 K and 2.5 K. The neutron wavelength was fixed at $\lambda = 2.4${\AA} to ensure greatest resolution from the thermal neutrons. 
Finally polycrystalline $\mu$SR measurements were carried out at the Swiss Muon Source of the Paul Scherrer Institut PSI, Villigen, using the $^3$He installation at GPD (MUE1). 
Transverse field (TF) experiments with a field of 50 Gauss (50 GTF) were performed as a function of temperature in order to systematically follow the magnetic behavior close to $T_N$.

\begin{figure}[!ht]
\begin{center}
\vspace{-2.0em} 
\includegraphics[width=0.8\linewidth]{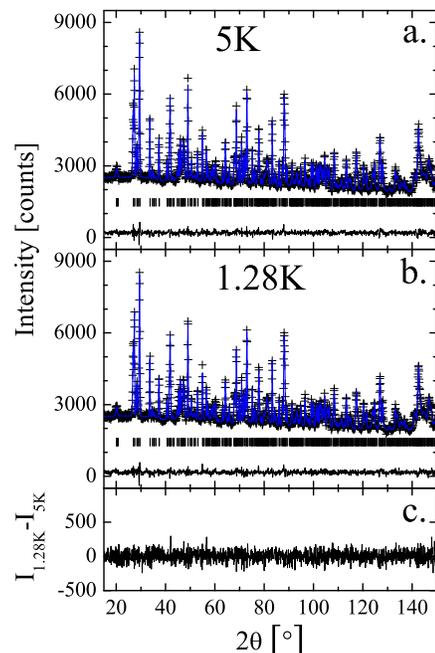}
\end{center}
\vspace{-3.5em} 
\caption{Powder neutron diffraction spectra of azurite at 5 K (a.) and in the AFM phase at 1.28 K (b.). Tics indicate structural Bragg peak positions, (+) represent experimental data. The solid line on the data shows the result of the refinement and the difference between refinement and fit is plotted at the bottom of each graph. In (c.) the difference between the spectra taken at 1.28 and 5 K is plotted.}
\vspace{-1.5em} 
\label{fig:fig1}
\end{figure}

The powder neutron diffraction results are summarised in Fig. \ref{fig:fig1}. These data were successfully refined using the {\scshape fullprof} program with the monoclinic crystal structure reported in the Refs. \cite{gattow,zigan,belokoneva} (blue solid line in Fig. \ref{fig:fig1}(a) and (b)). 
The lattice parameters found from the refinement of the 1.28K data set were $a$ = 5.002(1) \AA , $b$ = 5.825(1) \AA , $c$ = 10.342(1) \AA , and a monoclinic angle, $\beta = 92.21(1)^{\circ}$.
The error values $\chi^2 \approx 2$ and R$_{Bragg} \approx 4.8$ indicate the high quality of the fits.

The scattering intensity difference between the two data sets is neglibile, as displayed in \ref{fig:fig1}(c). Note that the noise amplitude is approximately 200 counts and may be attributed to incoherent scattering from the $^1$H in the sample. With such a poor signal to noise ratio, it is not surprising that the weak magnetic signal expected for the Cu$^{2+}$ ions in azurite were not observed above the background.
Thus the powder scattering pattern has revealed no evidence of magnetic order below $T_N$.

To gain more information, single crystal diffraction, which avoids the spherical integration inherent in powder diffraction, was used. 
The $a^* - b^*$-plane of azurite was explored below $T_N$ using the TAS diffractometer, E1.
No magnetic Bragg peaks were observed within this plane.

\begin{figure}[!ht]
\vspace{-1.0em}
\begin{center}
\includegraphics[width=0.8\linewidth]{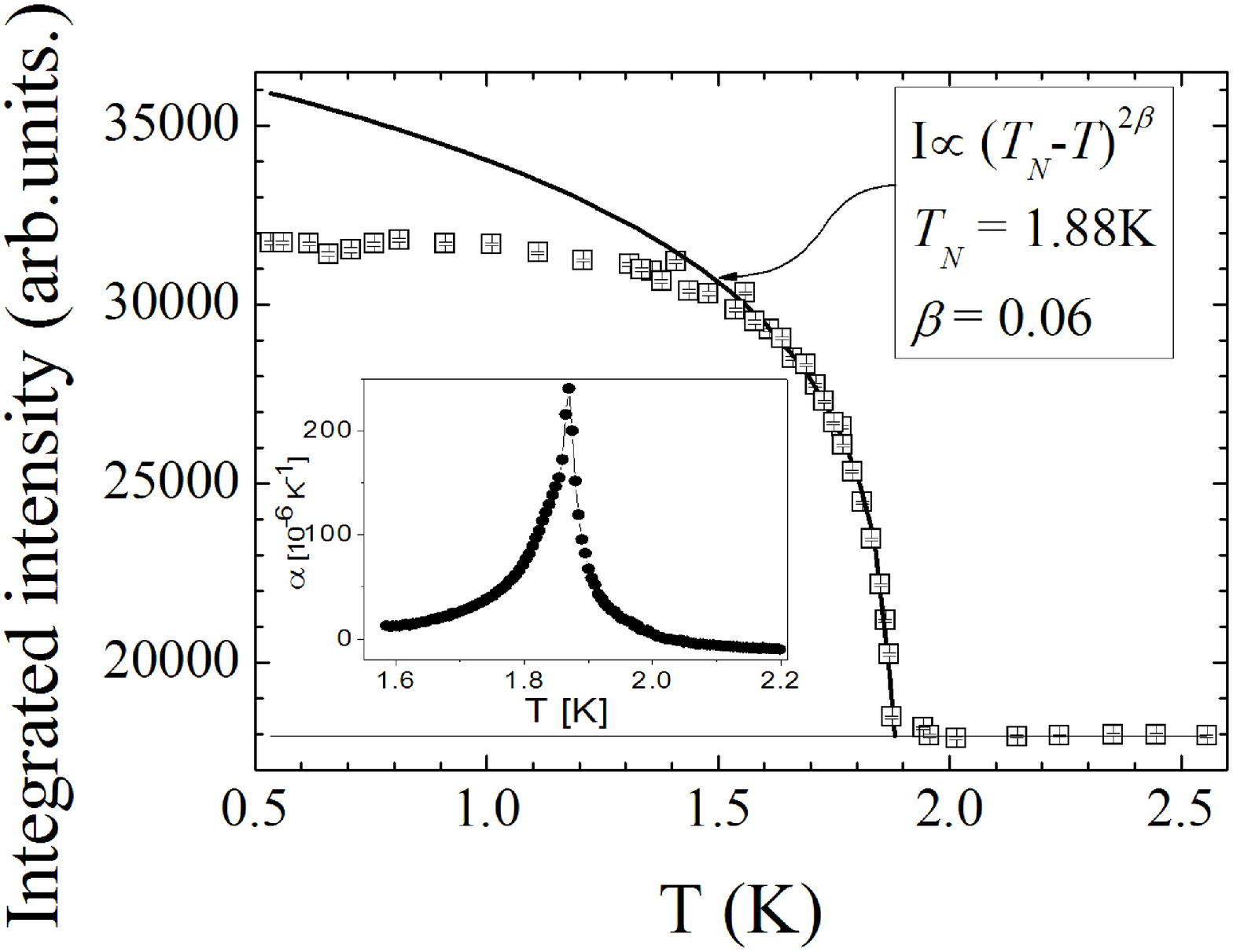}
\end{center}
\vspace{-2.5em}
\caption{The temperature dependence of the integrated intensity of the (120) Bragg peak of Cu$_3$(CO$_3$)$_2$(OH)$_2$, as obtained from single crystal neutron diffraction experiments. The inset displays the coefficient of uniaxial thermal expansion $\alpha$  along the b-axis as a function of temperature around $T_N$ = 1.88K.}
\vspace{-1.5em} 
\label{fig:fig3}
\end{figure}

The temperature evolution of the scattering intensity was observed over the quadrant in reciprocal space bounded by the points (000), (300), (030) and (330). 
For every Bragg peak which was measured in this quadrant we found a steep increase in intensity at $T_N$.
The temperature-dependence of the intensity of the (120) structural Bragg peak is illustrated in Fig. \ref{fig:fig3}.  
A rough calculation suggests that any magnetic scattering contribution from the copper atoms would be about $\frac{1}{40}^{th}$ of the nuclear scattering. Thus the additional $80\%$ intensity on the (120) peak cannot be magnetic in origin.

Fitting the temperature-dependence of the scattering intensity to the scaling expression, $(T_N - T)^{2\beta}$ we obtain the transition temperature $T_N = 1.88 K$ for our sample, in good agreement with the value obtained from bulk experiments.
However, for the critical exponent, $\beta$ we find a value of 0.06 which is extraordinarily small, again implying that the $T$-evolution of the additional Bragg peak intensity is not magnetic in origin.

In conjunction with the change of intensity of the structural Bragg peaks it was also found that the $b$-lattice parameter decreases significantly below $T_N$.  The inset of Fig. \ref{fig:fig3} exhibits the thermal expansion data measured along the $b$-axis in zero field. The uniaxial thermal expansion coefficient  $\alpha(T) = l^{-1} dl/dT$ reveals a $\lambda$-like anomaly at $T_N$ reflecting the character of the second order phase transition. The extraordinarily large $\alpha_b$ anomaly corresponds to a reduction in the $b$-axis lattice parameter upon cooling from 2 K to 1.6K of  $\Delta b/b = 22 \times 10^{-6}$ in agreement with the results given in Ref. \cite{Fabris}. 
It is thus shown that single crystal azurite exhibits lattice strain, coinciding with the onset of magnetic order at $T_N$.  However it is difficult to say if the structural strain influences the long range magnetic order, or if the magnetism forces the strain in the lattice.
A similar effect has been seen in the thermal expansion data of CuGeO$_3$ where sizable frustration of the AFM exchange was found to enhance the dimerisation of the spin-Peierls transition \cite{Buchner}.
Since the strain was not observed in the powder diffraction data, we can speculate that the difference in the grain size may play a role. The change in length as determined by the thermal expansion data indicates that the minimum grain size that would be affected by the observed lattice strain is around 10 $\mu$m.  As the average grain size in our powder sample is smaller than this, the observation of significant lattice strain is not expected in Fig. 1.
Thus a full low-temperature structural refinement is required for the single crystal sample to determine the combined influence of lattice strain and inner strain on the structure.
These strains and the change in atomic fractional coordinates influence the magnetic properties of azurite as they imply a subtle but important alteration of the superexchange pathways from the high temperature crystal structure. 

To confirm that magnetic order coincides with the structural distortion at $T \sim 1.9K$ we conducted powder $\mu$SR experiments.  For temperatures above about 3 K the sample signal consisted purely of a weakly damped Gaussian typical for a field distribution due to nearby nuclear moments. Meanwhile a strongly damped component appears for temperatures below 3K which is related to a magnetically ordered fraction. This can be traced in TF and also zero applied field (details will be reported in Ref. \cite{litterst}). As seen from Fig. 3 the normalised magnetic fraction, as deduced from the strongly damped contribution in 50 GTF, rises when lowering the temperature.  This is a gradual change down to about 2 K where a rapid increase follows in coincidence with $T_N$ claimed from specific heat\cite{forstat}. The change of damping below $T_N$ is rather smooth and supports a second order magnetic transition in azurite.  The increase in magnetic signal below 3K is reminicent of the maxima observed in susceptibility and heat capacity observed around 5K and 4K, respectively \cite{kikuchi2}. There it was related to the onset of correlations between monomers. We interpret our data below 3K with short-range order due to correlations between monomers induced via the coupling through the dimers which finally leads to the ordered phase below $T_N$. From these results we can conclude that the magnetic order becomes long-range below 1.8 K.  The apparent discrepancies in the onset temperatures for short range order, as obtained by muSR, specific heat and susceptibility, reflect the different time windows of these techniques.

\begin{figure}[!ht]
\vspace{-1.0em}
\begin{center}
\includegraphics[width=0.75\linewidth]{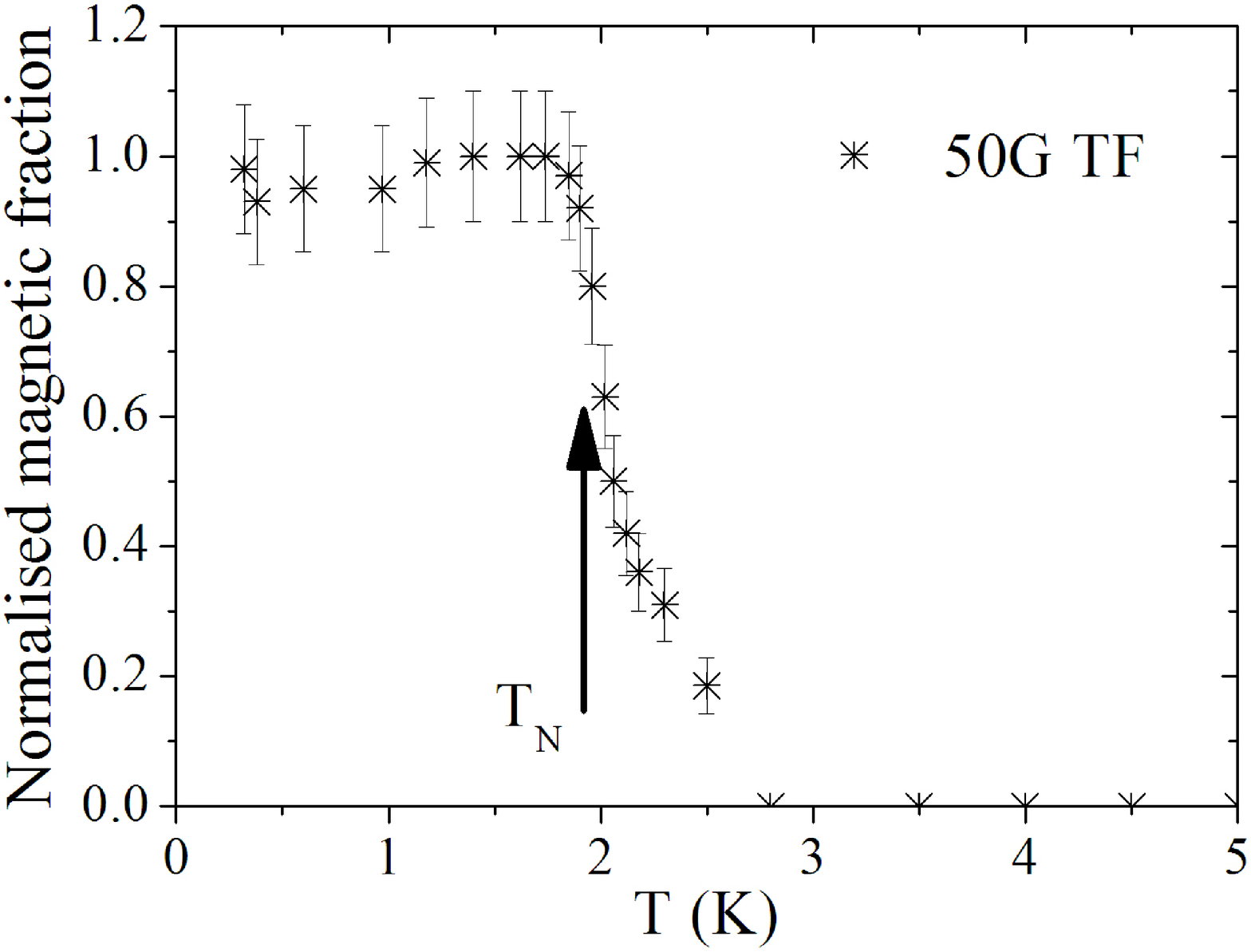}
\end{center}
\vspace{-2.5em}
\caption{Normalised magnetic fraction from 50 GTF (see text).}
\vspace{-0.75em}
\label{fig:figmusr}
\end{figure}  

To find magnetic ordering wavevectors of complex magnetic structures an effective method is 3D reciprocal space mapping. 
The E2 flat-cone diffractometer, equipped with 2D position sensitive detectors, accesses a large portion of $k$-space, with Miller indices from $-3< h <+3$ and $0< k <+3$ covered in the $a^*-b^*$plane and $-\frac{3}{10}< l <+\frac{3}{5}$ in the $c^*$-direction.

\begin{figure}[!ht]
\begin{center}
\includegraphics[width=1\linewidth]{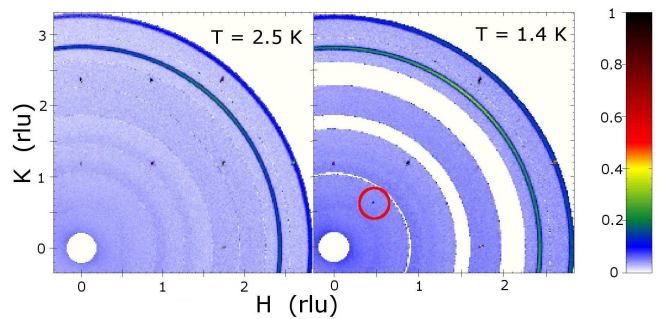}
\end{center}
\vspace{-2.0em}
\caption{(Colour online) Intensity distribution in reciprocal space taken from low temperature single crystal neutron diffraction of Cu$_3$(CO$_3$)$_2$(OH)$_2$. 
These plots show a cut along the $ab$-plane with all information out of the plane binned.  
White stripes indicate regions where no data was collected.  
The circle indicates an additional Bragg peak occurring at $\left(\frac{1}{2} \frac{1}{2} \frac{1}{2}\right)$ at temperatures below $T_N = 1.9$ K. 
The two high-intensity concentric circles visible in both diffraction patterns are powder diffraction lines from the sample mount.}
\vspace{-0.75em}
\label{fig:magpeak}
\end{figure}

On cooling below $T_N$, magnetic Bragg peaks were observed at the positions $\left(\pm\frac{1}{2} \frac{1}{2} \frac{1}{2}\right)$, $\left(\pm\frac{3}{2} \frac{5}{2} \frac{1}{2}\right)$ and $\left(\pm\frac{1}{2} \frac{3}{2} \frac{1}{2}\right)$.
The most intense of these is the $\left(\frac{1}{2} \frac{1}{2} \frac{1}{2}\right)$ peak, circled in red in figure \ref{fig:magpeak} implying that the magnetic peaks do not lie within the $ab$-plane.  This explains their absence from the TAS data and indicates an AFM structure with a magnetic unit cell double the size of the structural cell in each crystallographic direction. The other magnetic peaks, $\left(\pm\frac{3}{2} \frac{5}{2} \frac{1}{2}\right)$ and $\left(\pm\frac{1}{2} \frac{3}{2} \frac{1}{2}\right)$, are too weak to be observed with the intensity scale in figure \ref{fig:magpeak}.

Taking the symmetry of the crystallographic space group into account we can make some deductions about the magnetic structure of azurite. 
We consider only the monomer copper sites to be of importance in the magnetic ordering as the Cu$^{2+}$ dimer sites have only small local spin polarization and will not act independently \cite{aimo}.  The size of the spin moment on the dimer site is related to the staggered field of the neighbouring copper sites. The larger the spin gap of the dimer unit, the less influential the staggered field is in inducing a net spin moment on the dimer sites. Ohta and Rule \cite{ohta,rule} both found the size of the dimer spin gap to be large compared to the magnitude of the staggered field.
Based on the propagation vector of $\left(\frac{1}{2} \frac{1}{2} \frac{1}{2}\right)$ and using the SARAh-Representational Analysis program \cite{wills} we find two possible representations for the spin basis vectors, shown in table \ref{tab:rep}.  From the basis vectors it appears that the monomer sites on neighbouring chains along the \textit{c}-direction should be oriented perpendicular to each other.  Such a non--collinear configuration may suggest the presence of Dzyaloshinskii--Moriya interactions in the system.

Susceptibility measurements \cite{garber} indicate that the easy axis $x'$ is at approximately $55^{\circ}$ from the $c$-axis in the $a-c$ plane.
Note that in both of the representations, $\Gamma_{1}$ and $\Gamma_{2}$, the $x$ and $z$ components of the spins transform identically. We would therefore expect to be able to couple the $x$ and $z$ components of the spins such that the spins are constrained to lie in the $x'y$ plane. 
Further investigations will be required to obtain the precise orientation of the spin moments and the magnitude of the ordered magnetic moment.

\begin{table}[!ht]
	\centering
		\begin{tabular}{| l| c c |}
		\hline
		  IR & Cu(1) & Cu(2) \\ \hline
		  $\Gamma_{1}     $ & [1 1 1] & [$e^{i\pi/2}$ $e^{-i\pi/2}$ $e^{i\pi/2}$] \\ 
	          $\Gamma_{2}     $ & [1 1 1] & [$e^{-i\pi/2}$ $e^{i\pi/2}$ $e^{-i\pi/2}$] \\ \hline
		\end{tabular}
	\caption{The basis sets associated with the irreducible representations of the space group $\it P2_1/c$	and propagation vector $\bf{k}=$$\left(\frac{1}{2} \frac{1}{2} \frac{1}{2}\right)$. The basis sets give the relative phases of monomer spins at the 2 different Cu-sites Cu(1)=(000) and Cu(2)=$\left(\frac{1}{2} \frac{1}{2} \frac{1}{2}\right)$ in r.l.u.}
	\label{tab:rep}
\end{table}

It is curious that both magnetic ordering and structural distortion have been observed only in the single crystal and not in the powder neutron diffraction data. 
It is likely that the ordered magnetic moment in azurite is rather small which caused any magnetic scattering intensity to be obscured by incoherent scattering. It is also possible that the small grain size in the polycrystalline sample could have inhibited observation of the lattice strain.  
The single crystal TAS results have finally revealed the strong magnetostriction leading to a contraction of the $b$-axis below $T_N$. 
This structural distortion must be taken into account when estimating the microscopic magnetic exchanges with regards to the details of the superexchange pathways as it may imply a significant alteration of the Cu-O-Cu bond angles with respect to those of the high temperature structure.

Despite often being described as a good example of a 1D Heisenberg chain antiferromagnet, the low temperature properties of azurite appear to be more complex than first thought. 
The observation of a 3D ordered AFM phase implies that a simple 1D Heisenberg model is not sufficient to describe the system at low temperatures.  
Instead it seems necessary to consider further magnetic interactions such as interchain superexchange pathways and anisotropic exchange interactions such as Dzyaloshinskii–-Moriya couplings.  

The low temperature lattice parameters of azurite have been found from powder neutron diffraction results and provide essential new information for electronic structural calculations\cite{valenti}. 
The neutron results have also revealed the commensurate nature of the long range magnetic ordering in a single crystal. $\mu$SR results verify that magnetic order is also present in the polycrystalline material, while lattice strain in the single crystal is associated with the onset of the magnetically ordered phase. 

In conclusion, neutron scattering and $\mu$SR spectroscopy have revealed evidence of a magnetically ordered phase in azurite.  From these results a propogation vector of $\left(\frac{1}{2} \frac{1}{2} \frac{1}{2}\right)$ best describes the scattering of the ordered state which is reached via a second order phase transition.  Further to this, a large structural strain has been observed in azurite, coinciding with the transition to 3D AFM long-range order.

We would like to acknowledge partial support for muon experiments by European NMI3 programme of FP3 and experimental support at PSI by A. Maisuradze.  The authors also thank A. Honecker for fruitful discussions.  A.U.B would like to acknowledge financial support from the DFG grant DFG WO 1532/1-1.

\end{document}